%% file: 3SpinPaperMain.tex
\documentclass[a4paper,aps,prd,reprint,showpacs,showkeys,amsmath,amssymb,amsfonts,longbibliography,nofootinbib]{revtex4-1}

\usepackage{xcolor}
\xdefinecolor{mylinkcolor}{rgb}{0 0 0.5}

\usepackage[colorlinks=true,filecolor=mylinkcolor,citecolor=mylinkcolor,linkcolor=mylinkcolor,urlcolor=mylinkcolor,menucolor=mylinkcolor,bookmarksnumbered,bookmarksopen,bookmarksopenlevel=1]{hyperref}

\def\vct#1{\mathbf{#1}}

\def\nl{\\ & \quad}

\def\nla{\nonumber\\ &&}

\def\dim{d}

\def\pa{\partial}

\def\src#1{\mathcal{H}_{#1}^{\rm matter}}
\def\htt{h^{\rm TT}}

\def\pitt{\pi_{\rm TT}}

\def\gravthree{G}

\def\phib{{\bar{\phi}}}

\newcommand{\cInv}[1]{\left(c^{-1}\right)^{#1}}

\newcommand{\pitti}[1]{{\pitt^{#1}}} 
\newcommand{\pittis}[2]{{\pi_{(#1)\text{TT}}^{#2}}} 

\newcommand{\phis}[1]{{\phi_{(#1)}{}}} 
\newcommand{\srcs}[1]{\src{(#1)}} 
\newcommand{\srcsnontt}[1]{\src{(#1) \text{non-TT}}} 
\newcommand{\srcis}[2]{\src{#2 (#1)}} 
\newcommand{\phibs}[1]{{{\bar{\phi}}_{(#1)}{}}} 
\newcommand{\momls}[2]{{{\tilde{\pi}^{#2}_{(#1)}{}}}}
\newcommand{\vpots}[2]{{V^{#2}_{(#1)}}} 

\newcommand{\spin}[3]{\hat{S}_{#1\, (#2)(#3)}}
\def\canmom{P}
\newcommand{\mom}[2]{{\canmom}_{#1\,#2}}

\newcommand{\vmom}[1]{{\vct{\canmom}}_{#1}}
\newcommand{\vnxa}[1]{{\vct{n}}_{#1}}
\newcommand{\vspin}[1]{{\hat{\vct{S}}}_{#1}}
\newcommand{\vx}[1]{{\hat{\vct{z}}}_{#1}}
\newcommand{\xa}[2]{{\hat{z}}_{#1}^{#2}}
\newcommand{\dl}[1]{\delta_{#1}}

\newcommand{\sppr}[3]{{((#1\times#2)\,#3)}}
\newcommand{\crpr}[2]{{(#1\times#2)}}

\DeclareMathOperator{\Order}{\mathcal{O}}

\allowdisplaybreaks

\begin{document}

\title{Next-to-leading order spin-orbit and spin(a)-spin(b) Hamiltonians for 
$n$ gravitating spinning compact objects}

\author{Johannes Hartung}
\email{johannes.hartung@uni-jena.de}

\author{Jan Steinhoff}
\email{jan.steinhoff@uni-jena.de}

\affiliation{Theoretisch-Physikalisches Institut, Friedrich-Schiller-Universit\"at, Max-Wien-Platz\ 1, 07743 Jena, Germany}
\date{\today}

\begin{abstract}
We derive the post-Newtonian next-to-leading order conservative spin-orbit and
spin(a)-spin(b) gravitational interaction Hamiltonians for arbitrary many
compact objects. The spin-orbit Hamiltonian completes the knowledge
of Hamiltonians up to and including 2.5PN for the general relativistic three-body problem.
The new Hamiltonians include highly nontrivial three-body interactions,
in contrast to the leading order consisting of two-body interactions only.
This may be important for the study of effects like Kozai resonances in
mergers of black holes with binary black holes. The derivation
was done via two independent methods giving fully consistent results.
\end{abstract}


\maketitle
\input{3SpinPaperIntro.tex}

\input{3SpinPaperPre.tex}

\input{3SpinPaperCalc.tex}
\input{3SpinPaperRes.tex}
\input{3SpinPaperCon.tex}
\input{3SpinPaperApp.tex}

\acknowledgments
We thank G.\ Sch\"{a}fer, M.\ Tessmer, S.\ Hergt, and A.\ Gopakumar for many useful discussions.
Furthermore we thank P.\ Galaviz for pointing out a typo in our manuscript and hints on further references.
This work is partly funded by the Deutsche Forschungsgemeinschaft (DFG) through
the Research Training School GRK 1523 ``Quanten- und Gravitationsfelder'' and SFB/TR7 ``Gravitations\-wellen\-astronomie,''
as well as by the Deutsches Zentrum f\"ur Luft- und Raumfahrt (DLR) through ``LISA Germany.''

\renewcommand\bibAnnote[3]{}
\input{refs.bbl}

\end{document}

%% file: 3SpinPaperIntro.tex
\section{Introduction}
The gravitational interaction of $n$ compact objects is a fundamental
astrophysical problem. If one wants to tackle this problem within Einstein's
general relativity \cite{Einstein:1916}, then one generally has to resort to
numerical simulations, see, e.g., \cite{Lousto:Zlochower:2008,Campanelli:Lousto:Zlochower:2008,
Galaviz:Brugmann:Cao:2010}.
However, there exists a number of approximation methods.
One of the most successful approximation schemes is the post-Newtonian (PN)
approximation, a slow motion and wide separation approximation.
This allows an approximate solution of the field equations to some order,
leaving only ordinary differential equations for positions, momenta, and
spins of the compact objects. It is convenient to encode these equations
of motion in terms of a Lagrangian potential or a Hamiltonian. We will calculate
the PN approximate Hamiltonian via the canonical formalism of Arnowitt, Deser and Misner (ADM)
\cite{Arnowitt:Deser:Misner:1962}.

In the present paper we concentrate on spin contributions within
the post-Newtonian approximation. More specifically we will derive the conservative $n$-body
next-to-leading order (NLO) spin-orbit and spin(a)-spin(b) contributions,
where $a$ and $b$ label \emph{different} compact objects,
to the post-Newtonian Hamiltonian. These contributions were
already derived for the binary case $n=2$ in
\cite{Damour:Jaranowski:Schafer:2008:1,Steinhoff:Hergt:Schafer:2008:2}.
Other derivations can be found in \cite{Perrodin:2010,Porto:2010,Levi:2010,Porto:Rothstein:2008:1,Levi:2008}.
Next-to-leading order spin-orbit contributions to the equations
of motion for $n=2$ were first obtained in \cite{Tagoshi:Ohashi:Owen:2001}
and essentially confirmed in \cite{Faye:Blanchet:Buonanno:2006}.
The leading-order (LO) spin-orbit, spin(a)-spin(b), and spin(a)-spin(a)
contributions are well-known, see, e.g.,
\cite{Barker:OConnell:1975,DEath:1975,Barker:OConnell:1979,Thorne:Hartle:1985,Thorne:1980,Poisson:1998}.
For the next-to-leading order binary spin(a)-spin(a) interaction see
\cite{Steinhoff:Hergt:Schafer:2008:1,Hergt:Schafer:2008,Porto:Rothstein:2008:2,*Porto:Rothstein:2008:2:err,Steinhoff:Schafer:2009:1,Hergt:Steinhoff:Schafer:2010:1}.
Some binary Hamiltonians of even higher order in spin can be found in
\cite{Hergt:Schafer:2008:2,Hergt:Schafer:2008}.

The spin contributions to the dynamics have to be supplemented
by appropriate (i.e., sufficient within the approximation scheme)
point-mass\footnote{In terms of covariant multipole moments \cite{Dixon:1979},
for a point-mass all multipoles except the monopole are neglected.
The spin contributions in this paper arise from the covariant dipole
moment.} contributions. In general spin contributions can not
directly be compared to point-mass contributions as the spin is
a further expansion variable. However, for maximal spin, which is defined
by a ratio of spin to mass-squared corresponding to the extreme Kerr
solution, each power in spin is equivalent to half a post-Newtonian order.
Let us recall that for maximal spins the leading order spin-orbit Hamiltonian
is at 1.5PN order and the leading-order spin(a)-spin(b) one is at 2PN order,
while formally counted (i.e., without relating the spin variables to the PN counting)
both leading order Hamiltonians are at 1PN order. Similarly for maximal spins the next-to-leading
order Hamiltonians given in the present paper, formally having a post-Newtonian
order of 2, are comparable to 2.5PN point-mass contributions for
the spin-orbit case and to 3PN point-mass contributions for the
spin(a)-spin(b) case. The point-mass dynamics to 2PN was completed for
$n=3$ in \cite{Schafer:1987}, for corrections see \cite{Lousto:Nakano:2008},
and reduced to master integrals for arbitrary $n$ in \cite{Chu:2009}.
The spin-orbit Hamiltonian in the present paper thus completes the dynamics for maximal spins and $n=3$
to 2.5PN. (Notice that the dissipative 2.5PN point-mass dynamics
can trivially be extended from $n=2$ to arbitrary many objects, see, e.g.,
\cite{Jaranowski:Schafer:1997}.)
However, it should be emphasized that the results in the present
paper are valid for arbitrary $n$.
Also notice that spins close to maximal are astrophysically realistic,
see, e.g., \cite{McClintock:Narayan:Gou:Penna:Steiner:2009}.

Until the point-mass contributions to the post-Newtonian approximation are not
pushed to a higher number of objects $n$, the most useful application
of the Hamiltonians given in the present paper is the investigation
of the \emph{three}-body problem with rapidly rotating objects in general
relativity. This ideally fits to numerical investigations as in
\cite{Lousto:Zlochower:2008,Campanelli:Lousto:Zlochower:2008,Galaviz:Brugmann:Cao:2010}, 
which are accurate beyond the applicability
of the post-Newtonian approximation but require much more computational
power. An important astrophysical application is the investigation of
hierarchical triplets. The Hamiltonians provided in this paper allow
an accurate treatment of, e.g., Kozai resonances
\cite{Kozai:1962,Ford:Kozinsky:Rasio:2000,*Ford:Kozinsky:Rasio:2000:err}
in mergers of a black hole with a black hole binary when one or
several of these black holes are rapidly rotating. One may also try to
find stable solutions, such as the periodic ones for non-spinning objects
given in \cite{Moore:1993,Imai:Chiba:Asada:2007,Lousto:Nakano:2008}.
Further the three-body problem is always interesting for the study of chaotic behavior.
To foster such application the derived Hamiltonians for three compact objects are provided as
Mathematica source files \cite{sourcefiles}.

The paper is organized as follows. In Sec. \ref{sec:ADM}, we provide a short introduction to the ADM formalism. 
After this, in Sec. \ref{sec:Calc}, a few details (namely constraint expansions, integration by parts and 
three-body integrals) of the calculation will be explained. In Sec. \ref{sec:Results} the results for 
the Hamiltonians and checks (and for the readers convenience the appropriate center of mass vectors) will be provided. 
Last but not least, there will be some conclusions and further tasks given in Sec. \ref{sec:Conclusions}.

The signature of spacetime is $+2$ in the present paper.
Since the PN formalism is a perturbation theory around a flat Euclidean background it does not matter 
in principle whether the spatial indices in some tensor expressions are upper or lower ones (although the index position 
is important for the definition of some quantities). These indices are denoted by small Latin letters from the 
middle of the alphabet ($i,j,k,\dots$) and running from $1$ to $3$. Greek indices $(\mu,\nu,\dots)$ are 4-dimensional
indices running from $0$ to $3$. Object labels are denoted by small Latin letters from the beginning of the 
alphabet ($a,b,c,\dots$). In this paper we sum over all double indices (Einstein summation convention) except 
object indices. Sums over object labels are explicitly written in the expressions. Vectors are denoted by 
boldface letters and the scalar product between two vectors $\vct{a}$ and $\vct{b}$ is denoted by 
$(\vct{a}\vct{b}) \equiv (\vct{a} \cdot \vct{b})$. 
Furthermore, the speed of light $c$ is set equal to $1$ and if there is a $\cInv{}$ appearing it is just a bookkeeping parameter to 
get the correct post-Newtonian order of quantities or expressions.
The Newton gravitational constant is denoted by $\gravthree$
and we did not use a special convention for it. The reader may set $\gravthree$ to a desired value in our expressions.

%% file: 3SpinPaperPre.tex
\section{The ADM Formalism}\label{sec:ADM}
In the present paper, we will utilize the ADM canonical formalism
after gauge fixing \cite{Arnowitt:Deser:Misner:1962}, see also \cite{Regge:Teitelboim:1974,DeWitt:1967}.
At this stage the constraints of the gravitational
field are solved (approximately in our case). The Hamiltonian is then
given by the ADM energy expressed in terms of certain canonical variables.
The ADM formalism has shown to be valuable for calculating the
conservative dynamics within the post-Newtonian and post-Minkowskian approximations, see, e.g.,
\cite{Jaranowski:Schafer:1998,Damour:Jaranowski:Schafer:2001,Ledvinka:Schafer:Bicak:2008}.

The constraints of the gravitational field can be written as
\begin{gather}
\frac{1}{16\pi\gravthree \sqrt{\gamma}}\left[\gamma\text{R}+\frac{1}{2}\left(\gamma_{ij}\pi^{ij}\right)^2 - \gamma_{ij}\gamma_{k\ell}\pi^{ik}\pi^{j\ell}\right] = \src{} , \\
-\frac{1}{8\pi\gravthree}\gamma_{ij}\pi^{jk}_{\quad;k} = \src{i} \,,
\end{gather}
with the definitions
\begin{align}
\pi^{ij} &= - \sqrt{\gamma} (\gamma^{ik}\gamma^{jl} - \gamma^{ij}\gamma^{kl})K_{kl} \,, \\
\src{} &= \sqrt{\gamma}T_{\mu\nu} n^{\mu}n^{\nu} \,, \\
\src{i} &= - \sqrt{\gamma}T_{i \nu} n^{\nu} \,,
\end{align}
and arise as certain projections of the Einstein equations with respect to a timelike
unit 4-vector $n_{\mu}$ with components $n_{\mu} = (-N, 0, 0, 0)$ or $n^{\mu} = (1, -N^i) / N$.
Here $\gamma_{ij}$ is the induced three-dimensional metric of the hypersurfaces orthogonal to $n_{\mu}$,
$\gamma$ its determinant, $\text{R}$ the three-dimensional Ricci scalar, $K_{ij}$ the extrinsic curvature,
$N$ the lapse function, $N^i$ the shift vector, $ \sqrt{\gamma}T_{\mu\nu}$ the stress-energy
tensor density of the matter system, and $;$ denotes the three-dimensional covariant derivative.
Partial coordinate derivatives $\pa_i$ are also indicated by a comma. 
For the extrinsic curvature $K_{ij}$ we used the ADM sign convention, i.e.,
$2 N K_{ij} = - \gamma_{ij,0} + N_{i;j} + N_{j;i}$.

In the ADM transverse-traceless (ADMTT) gauge defined by
\begin{align}
3\gamma_{ij,j} - \gamma_{jj,i} &= 0 \,, \label{ADMTTg} \\
\pi^{ii} &= 0 \,, \label{ADMTTpi}
\end{align}
which will be used throughout this paper, one has the decompositions
\begin{align}
	\gamma_{ij} &= \left( 1 + \frac{\phi}{8} \right)^4 \delta_{ij} + \htt_{ij} \,,
		\label{gdecomp} \\
	\pi^{ij} &= \pitti{ij} + \tilde{\pi}^{ij} \,,\label{eq:pidecomp}
\end{align}
where $\htt_{ij}$ and $\pitti{ij}$ are symmetric and transverse-traceless,
e.g, $\htt_{ij}=\htt_{ji}$, $\htt_{ii} = \htt_{ij,j}=0$.
Notice that the form of the trace term in (\ref{gdecomp})
is adapted to the Schwarzschild metric in isotropic coordinates,
with obvious advantages for perturbative expansions.
For our convenience we introduced a rescaled $\phib \equiv \phi/8$, which is useful 
later in the expansion of the constraint equations. 
The longitudinal part, $\tilde{\pi}^{ij}$, of $\pi^{ij}$ in Eq. (\ref{eq:pidecomp})
can be written in two equivalent forms, either in terms of $\tilde{\pi}^i$ (which contains
an inverse Laplacian $\Delta^{-1}$),
\begin{align}
\tilde{\pi}^{ij}
	&= \tilde{\pi}^i_{, j} + \tilde{\pi}^j_{, i}
	- \frac{1}{2} \delta_{ij} \tilde{\pi}^k_{, k}
	- \frac{1}{2} \Delta^{-1} \tilde{\pi}^k_{, ijk} \,,
\end{align}
or in terms of $V^i$ (which contains no inverse Laplacian),
\begin{align}
\tilde{\pi}^{ij} &= V^i_{, j}
	+ V^j_{, i} - \frac{2}{3} \delta_{ij} V^k_{, k} \,.
\end{align}
The two vector potentials $\tilde{\pi}^i$ and $V^i$ are related by
\begin{align}
V^i &= \left( \delta_{ij} - \frac{1}{4} \partial_i \partial_j \Delta^{-1} \right) \tilde{\pi}^j \,, 
\end{align}
and can be obtained as solutions of the momentum constraint via
\begin{align}
\tilde{\pi}^i &= \Delta^{-1} \pi^{ij}_{,j}
	= \Delta^{-1} \tilde{\pi}^{ij}_{,j} \,,
\end{align}
cf.\ Eq.\ (\ref{momcexpand}).
The transverse-traceless(TT) part of $\pi$, $\pitti{ij}$, is given by
\begin{align}
 \pitti{ij} &= \delta^{\text{TT}ij}_{kl} \pi^{kl} \,,
\end{align}
with the partial space-coordinate derivatives $\pa_i$ and
the transverse-traceless projector
\begin{equation}\label{TTproj}
\begin{split}
\delta^{\text{TT}kl}_{ij} &= \tfrac{1}{2} [(\delta_{il}-\Delta^{-1}\pa_{i}\pa_{l})(\delta_{jk}-\Delta^{-1}\pa_{j}\pa_{k}) \nl
+(\delta_{ik}-\Delta^{-1}\pa_{i}\pa_{k})(\delta_{jl}-\Delta^{-1}\pa_{j}\pa_{l}) \nl -(\delta_{kl}-\Delta^{-1}\pa_{k}\pa_{l})(\delta_{ij}-\Delta^{-1}\pa_{i}\pa_{j})] \,.
\end{split}
\end{equation}

Now the four field constraints can be solved for the four variables $\phi$ and $\tilde{\pi}^i$
in terms of $h^{\text{TT}}_{ij}$, $\pi^{ij}_{\text{TT}}$
and matter variables, which enter through the stress-energy tensor via
the source terms $\mathcal{H}^{\rm matter}$ and $\mathcal{H}^{\rm matter}_i$.
An analytic solution for $\phi$ and $\tilde{\pi}^i$, however,
can in general only be given in some approximation scheme.
Finally, the ADM Hamiltonian $H_{\text{ADM}}$ reads
\begin{equation}\label{HADM}
H_{\text{ADM}} = - \frac{1}{16\pi\gravthree} \int \text{d}^3x \, \Delta
	\phi \,. 
\end{equation}
This is the ADM energy expressed in terms of the canonical variables.
The canonical matter variables are introduced in Sec.\ \ref{source} below.
The canonical field variables are $h^{\text{TT}}_{ij}$ and $\pi^{ij}_{\text{TT}}$
here, with the Poisson brackets
\begin{equation}
\{ \htt_{ij}({\bf x}), \pitti{kl}({\bf x}') \}
	= 16\pi\gravthree\, \delta^{\text{TT}kl}_{ij}\delta({\bf x} - {\bf x}') \,.
\end{equation}
Notice that beyond the post-Newtonian order considered here
spin corrections to the canonical field momentum are needed
\cite{Steinhoff:Schafer:2009:2,Steinhoff:Wang:2009}.

%% file: 3SpinPaperCalc.tex
\section{Calculation}\label{sec:Calc}
The field and source expansions starting at their leading order are given by
\begin{subequations}
\begin{eqnarray}
 \phi & = & \phis{2} + \phis{4} + \phis{6} + \phis{8} + \dots \,,\\
 \tilde{\pi}^{ij} & = & \momls{3}{ij} + \momls{5}{ij} + \dots \,,\\
 \src{} & = & \srcs{2} + \srcs{4} \nonumber\\
&& + \srcs{6} + \srcs{8} + \dots \,, \\
 \src{i} & = & \srcis{3}{i} + \srcis{5}{i} + \dots \,,
\end{eqnarray}
\end{subequations} 
where the subscript in round brackets denotes the $\cInv{}$ order. The $\htt_{ij}$ field only occurs in leading order, namely $\cInv{4}$. Since the TT field momentum is related to time derivatives of $\htt_{ij}$ the 
leading order of $\pitti{ij}$ is $\cInv{5}$. The mass $m_a$, canonical matter momentum $\vmom{a}$, and spin variables $\vspin{a}$ are formally counted as $m_a \sim \Order{\cInv{2}}$, $\vmom{a} \sim \Order{\cInv{3}}$, and $\vspin{a} \sim \Order{\cInv{3}}$ for dimensional reasons only (remember that for maximal spins one would have $\vspin{a} \sim \Order{\cInv{4}}$ instead).
This counting comes from the fact that after setting $c=\gravthree=1$ we require all quantities to be in units of length. Let us introduce
symbols with a bar over them being the quantities in SI units and the other symbols the quantities in units of length, then 
it holds $m_a = \tfrac{\gravthree}{c^2} \bar{m}_a$ for the mass, $t = c \bar{t}$ for the time,
$\vmom{a} = \tfrac{\gravthree}{c^3} \bar{\vct{\canmom}}_{a}$ for the linear momentum,
and similar for the spin variables.
So the order counting comes from the $c$ powers inserted to reconstruct the SI units.
It should be noted that these counting rules will in general not give correct \emph{absolute} orders in $c$ if the SI units of the
final expression are not taken into account. However, \emph{relative} orders are always meaningful, which is all that is relevant
for perturbative expansions. Further notice that different counting rules are obtained if one assumes that all quantities are
expressed in terms of mass units instead of length units when setting $c=\gravthree=1$, which is also often used in the literature.

\subsection{Constraint expansions}
The Hamilton constraint expansion is given by
\begin{widetext}
\begin{subequations}\label{ham}
\begin{eqnarray}
 -\frac{1}{16 \pi \gravthree} \Delta \phis{2} & = & \srcs{2}\,,\\
 -\frac{1}{16 \pi \gravthree} \Delta \phis{4} & = & \srcs{4} - \phibs{2} \srcs{2}\,,\\
 -\frac{1}{16 \pi \gravthree} \Delta \phis{6} & = & \srcs{6} - \phibs{2} \srcs{4} + (-\phibs{4} + \phibs{2}^2) \srcs{2} - \frac{1}{16 \pi \gravthree}\biggl(- (\momls{3}{ij})^2 + 4 (\phibs{2} \htt_{(4)ij})_{,ij}\biggr)\,,\label{eq:HamConstr1PN}\\
 -\frac{1}{16 \pi \gravthree} \Delta \phis{8} & = & \srcs{8} - \phibs{2} \srcs{6}  + (-\phibs{4} + \phibs{2}^2)\srcs{4} + ( - \phibs{6} + 2 \phibs{2} \phibs{4} -\phibs{2}^3) \srcs{2}\nonumber\\
&& - \frac{1}{16 \pi \gravthree } \biggl(- \phibs{2} (\momls{3}{ij})^2 - 2 \momls{3}{ij} \momls{5}{ij} - 2 \momls{3}{ij} \pittis{5}{ij} + 4 \htt_{(4)ij} \phibs{2}_{,i}\phibs{2}_{,j} - \frac{1}{4} (\htt_{(4)ij,k})^2 \biggr)\nonumber\\
&&+\text{(td)}\,,
\end{eqnarray}
\end{subequations}
where $\Delta=\pa_i\pa_i$.
To $\cInv{5}$ order, the momentum constraint can be expanded via
\begin{eqnarray}\label{momcexpand}
 \momls{3}{ij}{}_{,j} = - 8 \pi \gravthree \srcis{3}{i}\,, \qquad
 \momls{5}{ij}{}_{,j} = - 8 \pi \gravthree \srcis{5}{i} - (4 \momls{3}{ij} \phibs{2})_{,j} \,. 
\end{eqnarray}


\subsection{Source expansion\label{source}}

The source of the field constraints $\mathcal{H}^{\rm matter}$ and $\mathcal{H}^{\rm matter}_i$
in terms of canonical variables were derived for spinning objects to linear order in the single
spin variables and to the post-Newtonian order required in this
paper in \cite{Steinhoff:Schafer:Hergt:2008}. Higher post-Newtonian orders
were treated in \cite{Steinhoff:Wang:2009} and the formalism was worked
out to all orders in \cite{Steinhoff:Schafer:2009:2}.
The expansion of the source into powers of $1/c$ reads
\begin{subequations}
\begin{eqnarray}
 \srcs{2} & = & \sum_a m_a \dl{a}\,,\\
 \srcs{4} & = & \sum_a \biggl[\frac{\vct{\canmom}^2_a}{2 m_a} \dl{a} + \frac{1}{2 m_a} \mom{a}{i} \spin{a}{i}{j} \dl{a}{}_{,j}\biggr]\,,\\
 \srcs{6} & = & \sum_a \biggl[-\frac{(\vct{\canmom}^2_a)^2}{8 m_a^3} \dl{a} - 2 \frac{\vct{\canmom}_a^2}{m_a} \phibs{2}\dl{a} + 2 \frac{\mom{a}{i}}{m_a}\spin{a}{i}{j} \phibs{2}_{,j}\dl{a}
 - \frac{\vct{\canmom}_a^2}{8 m_a^3} \mom{a}{i}\spin{a}{i}{j} \dl{a}{}_{,j} - 2 \frac{\mom{a}{i}}{m_a} \spin{a}{i}{j} (\phibs{2} \dl{a})_{,j}\biggr] , \hspace{0.7cm} \\
\srcs{8} & = & \sum_a \biggl[\frac{(\vct{\canmom}^2_a)^3}{16 m_a^5} \dl{a} +  \frac{(\vct{\canmom}_a^2)^2}{m_a^3}\phibs{2}\dl{a} + 5 \frac{\vct{\canmom}_a^2}{m_a} \phibs{2}^2 \dl{a} - 2 \frac{\vct{\canmom}_a^2}{m_a} \phibs{4} \dl{a}
- \frac{1}{2 m_a} \mom{a}{i} \mom{a}{j} \htt_{(4)ij}\dl{a} - \frac{\vct{\canmom}_a^2}{m_a^3} \mom{a}{i}\spin{a}{i}{j}  \phibs{2}_{,j}\dl{a}\nonumber \\ && - 10 \frac{\mom{a}{i}}{m_a} \spin{a}{i}{j} \phibs{2}\phibs{2}_{,j} \dl{a}
 + 2 \frac{\mom{a}{i}}{m_a} \spin{a}{i}{j} \phibs{4}_{,j} \dl{a} + \frac{1}{2 m_a} \mom{a}{i} \spin{a}{j}{k} \htt_{(4)ij,k} \dl{a}\biggr]+\text{(td)}\,,
\end{eqnarray}
\end{subequations}
\end{widetext}
in the case of the Hamilton constraint sources
and for the momentum constraint sources one obtains
\begin{subequations}
 \begin{align}
   \srcis{3}{i} & = \sum_a \biggl[\mom{a}{i} \dl{a} + \frac{1}{2} (\spin{a}{i}{j} \dl{a})_{,j}\biggr]\,,\\
 \srcis{5}{i} & =  \frac{1}{2} \sum_a \biggl[ -\frac{\mom{a}{k}}{2 m_a^2} (\mom{a}{j}\spin{a}{i}{k} + \mom{a}{i} \spin{a}{j}{k}) \dl{a}\biggr]_{,j}\,.
 \end{align}
\end{subequations}
Here $\mom{a}{i}$ are the matter canonical momenta, $\dl{a} = \delta(x^i - \hat{z}_a^i)$
with $\hat{z}_a^i$ the canonical position variable, and $\spin{a}{i}{j} = - \spin{a}{j}{i}$ is the
canonical spin tensor. The latter is related to the spin vector $\hat{S}_{a(i)}$ by
$\spin{a}{i}{j} = \epsilon_{ijk} \hat{S}_{a(k)}$ where $\epsilon_{ijk}$ is the Levi-Civita symbol.
These variables have the canonical Poisson brackets
\begin{align}
\{ \hat{z}_a^i , \mom{a}{j} \} &= \delta_{ij} \,, \\
\{ \hat{S}_{a(i)} , \hat{S}_{a(j)} \} &= \epsilon_{ijk} \hat{S}_{a(k)} \,, \label{spinpb}
\end{align}
all other zero. Notice that the spin-length $S_a^2 \equiv \hat{S}_{a(i)} \hat{S}_{a(i)}$ is
constant as all its Poisson brackets vanish. Therefore the spin has only
two dynamical degrees of freedom. For some applications it is useful to work in
a basis of phase space which makes this explicit, especially for investigations
regarding chaos \cite{Wu:Xie:2010}. If one parametrizes the spin vectors as
\begin{equation}
( \hat{S}_{a(i)} ) = S_a \left( \begin{array}{c}
	\sin\theta_a \cos\phi_a \\
	\sin\theta_a \sin\phi_a \\
	\cos\theta_a
\end{array} \right) \,,
\end{equation}
then possible canonical variables are the pairs $\phi_a$ and $\hat{S}_{a(3)}=S_a \cos\theta_a$ with
\begin{equation}\label{spinpb2}
\{ \phi_a , \hat{S}_{a(3)} \} = 1 \,,
\end{equation}
all other zero, see \cite{Bel:Martin:1980} and also \cite{Damour:Jaranowski:Schafer:2008:1}.
However, this introduces square roots as
\begin{equation}\label{spinpar}
( \hat{S}_{a(i)} ) = \left( \begin{array}{c}
	\sqrt{1 - \hat{S}_{a(3)}^2} \cos\phi_a \\
	\sqrt{1 - \hat{S}_{a(3)}^2} \sin\phi_a \\
	\hat{S}_{a(3)}
\end{array} \right) \,.
\end{equation}
It is straightforward to check that (\ref{spinpar}) and (\ref{spinpb2})
lead to (\ref{spinpb}).

\subsection{Integration by parts}
The post-Newtonian expanded ADM Hamiltonian results according to (\ref{HADM}) from
an integral over the right-hand side of (\ref{ham}). However, this integral
can be greatly simplified.

First of all one can get rid of the $\momls{3}{ij}\pittis{5}{ij}$ term via integration by parts, since $\pittis{5}{ij}$ is divergence free and one can rewrite $\momls{3}{ij}$ in terms of derivatives of the $V^i_{(3)}$ vector potential.
Furthermore one can eliminate $\phibs{6}\srcs{2}$ via integration by parts and using Eq. (\ref{eq:HamConstr1PN}) in the system of constraint equations. One can eliminate the $\momls{5}{ij}$ via rewriting $\momls{3}{ij}$ in terms of $V^i_{(3)}$ derivatives as well and gets a source type term and another $\phibs{2}(\momls{3}{ij})^2$ contribution.

After these integrations by parts one can change from a Hamiltonian in the TT degrees of freedom to a so called Routhian, a Hamiltonian in the particle degrees of freedom and a Lagrangian in the propagating field degrees of freedom \cite{Jaranowski:Schafer:1998}. The TT degrees of freedom are then eliminated from the Routhian by inserting their approximate solution. (The reason for this is that one can insert equations of motion for the time derivatives of the particle variables appearing in the velocities of the TT degrees of freedom, which corresponds to a coordinate transformation only \cite{Schafer:1984}.) The Hamiltonian resulting from this procedure is given by
\begin{eqnarray}
 H_{\text{2PN}} & = & H^\text{matter}_{\text{2PN}} + H^\text{TT}_{\text{2PN}}\,,
\end{eqnarray}
where
\begin{align}
\begin{split}
H^{\text{matter}}_{\text{2PN}} &= \int \text{d}^3 x \biggl[\srcsnontt{8} - 2 \phibs{2} \srcs{6} \nl + (-\phibs{4} +2 \phibs{2}^2)\srcs{4} \nl + ( 3 \phibs{2} \phibs{4} - 2 \phibs{2}^3) \srcs{2} \nl + 2\srcis{5}{i} \vpots{3}{i}- \frac{1}{16 \pi \gravthree } \,8 \phibs{2} (\momls{3}{ij})^2\biggr] \,,
\end{split}\\
H^{\text{TT}}_{2\text{PN}} &= \frac{1}{16\pi\gravthree} \int \text{d}^3 x \, \frac{1}{2} B_{(4)ij} h^{\text{TT}}_{(4)i j} \,,\\
B_{(4)ij} &= 16\pi\gravthree \frac{\delta \left(
		\int{ \text{d}^3 x \, \srcs{8}  } \right)}
		{\delta h^{\text{TT}}_{i j}}
	- 8 \phibs{2}{}_{, i} \phibs{2}{}_{, j} \,, \\
h^{\text{TT}}_{(4)i j} &= 2 \delta^{\text{TT}kl}_{ij} \Delta^{-1} B_{(4)kl} \,,
\end{align}
where $\Delta^{-1}$ is the inverse Laplacian for usual boundary conditions.
It is possible to rewrite $H^{\text{TT}}_{2\text{PN}}$ into a form where no point-mass part of $\htt_{(4)ij}$ is needed, see \cite{Steinhoff:Schafer:Hergt:2008} for details.

\subsection{Three-body integrals}
Since there are at most three fields appearing (and no field which is generated by more than one body) in the integral for the Hamiltonian at 2PN spin-orbit and spin(a)-spin(b) level, namely
\begin{eqnarray}
 H^{\text{matter,SO}}_{\phi(2)\pi(3)^2}&=& -\frac{1}{16 \pi \gravthree } \int\text{d}^3 x\,16 
\phibs{2} \momls{3}{ij}_{\text{PP}}\momls{3}{ij}_{\text{S}}\,,\nonumber\\
\\
 H^{\text{TT-part,SO}}_{\partial \phi(2)^2\htt{}{(4)}}&=&
 -\frac{1}{16\pi \gravthree} \int\text{d}^3 x\,
8\phibs{2}_{,i}\phibs{2}_{,j}\htt_{(4)\text{S}\,ij}\,,\nonumber\\
\\
 H^{\text{matter,SS}}_{\phi(2)\pi(3)^2}&=&-\frac{1}{16 \pi \gravthree } \int\text{d}^3 x\,8 
\phibs{2} (\momls{3}{ij}_{\text{S}})^2\,,
\end{eqnarray}
we will not get any integrals where a higher number of compact objects is involved.
The abbreviations PP, SO, S, SS, and S$^2$ (some of them first appear later) stand for point-mass part 
(or point-particle part), spin-orbit part, spin part of the field, spin(a)-spin(b) part, and 
spin(a)-spin(a) part, respectively.
(There is another three-body integral generated by two fields and one delta source in the integrand, 
$H_{\delta-\text{type}}^{\text{matter,SO}}$, for which we do not need the calculation procedures 
mentioned here and which is given in the results later.) So it is sufficient to calculate only 
three-body integrals for the $n$-body 2PN spin-orbit and spin(a)-spin(b) contribution to the Hamiltonian.
We refer to integrals of the type mentioned above as three-body integrals, because 
they describe an interaction between three different position variables (if the object positions are 
not distinct, one will get a two-body or one-body integral).

The only three-body integrals appearing here are well-known and were already solved in three dimensions, namely
\begin{subequations}
\begin{eqnarray}
 \int \text{d}^3 x \frac{1}{r_a r_b r_c} & = & -4\pi \left.\Delta^{-1} \frac{1}{r_a r_b}\right|_{\vct{x} = \vx{c}}\nonumber\\
 &=& -4\pi \ln s_{abc}\,,\label{eq:logintegral}\\
 \int \text{d}^3 x \frac{r_b}{r_a  r_c} & = & -4\pi \left.\Delta^{-1} \frac{r_b}{r_a }\right|_{\vct{x} = \vx{c}}\nonumber\\
& = &-4\pi\biggl\{\frac{1}{18}(3 r_{ac} r_{bc} + 3 r_{ac} r_{ab} - 3 r_{ab} r_{bc}\nonumber\\
&&\quad-r_{bc}^2 - r_{ab}^2  + r_{ac}^2) \nonumber\\
&&+ \frac{1}{6} (r_{bc}^2 + r_{ab}^2 - r_{ac}^2)\ln s_{abc}\biggr\}\,,\label{eq:x2logintegral}
\end{eqnarray}
\end{subequations}
where $s_{abc} = r_{ab} + r_{ac} + r_{bc}$, $r_a = |\vct{x}-\vx{a}|$ and $r_{ab} = |\vx{a}-\vx{b}|$. 
The first integral was solved in \cite[Eq. (82,33)]{Fock:1960} and \cite{Ohta:Okamura:Kimura:Hiida:1973}, and 
the second one in \cite{Jaranowski:Schafer:1998}. Note that the integrals on the left-hand side of 
(\ref{eq:logintegral}) and (\ref{eq:x2logintegral}) are only formal expressions since they are divergent. 
Their (regularized) solutions on the right-hand side are not unique and have to be fixed by certain consistency conditions, 
e.g., that Laplacians operating on different particle coordinates give certain functions. Further 
the integrals in the form given above are auxiliary functions, only their derivatives, which in fact are convergent,
enter the physical expressions. See discussion in, e.g., \cite{Jaranowski:Schafer:1998} for further details.

After inserting these integrals, it was necessary to rewrite derivatives with respect to $\vct{x}$ 
into derivatives with respect to, e.g., $\vx{a}$ to pull them out of the integral.
(Derivatives with respect to components of particle coordinates $\vx{a}$ are denoted by $\partial_i^{(a)}$.) 
So in principle these parts of the Hamiltonian can be calculated by integrating appearing three-body 
integrals and afterwards differentiate them by the different particle coordinates three, four, or five times depending 
on the appropriate part of the Hamiltonian. The parts of the Hamiltonians which were calculated by the algorithm mentioned above are given by
\begin{eqnarray}
H^{\text{matter,SO}}_{\phi(2)\pi(3)^2}&=&2\gravthree^2 \sum_{a} \sum_{b\ne a}\sum_{c\ne a,b} m_a \spin{c}{\ell}{i}\nonumber\\
&&\biggl\{\biggl[4  \mom{b}{i} \partial_j^{(b)} \partial_j^{(c)} \partial_\ell^{(c)}\nonumber\\
&&\quad +4\mom{b}{j}\partial_i^{(b)} \partial_j^{(c)} \partial_\ell^{(c)} \biggr]\ln s_{abc}\nonumber\\
&&- \mom{b}{k}\partial_i^{(b)}\partial_j^{(b)} \partial_k^{(b)} \partial_j^{(c)} \partial_\ell^{(c)}\nonumber\\
&&\biggl[\frac{1}{18}(3 r_{ac} r_{bc} + 3 r_{ac} r_{ab} - 3r_{ab} r_{bc}\nonumber\\
&& - r_{bc}^2- r_{ab}^2 + r_{ac}^2) \nonumber\\
&&+ \frac{1}{6} (r_{bc}^2 + r_{ab}^2 - r_{ac}^2)\ln s_{abc}\biggr]\biggl\}\,,\\ 
 H^{\text{TT-part,SO}}_{\partial \phi(2)^2\htt{}{(4)}} & = & -2\gravthree^2 \sum_{a}\sum_{b\ne a}\sum_{c\ne a,b} \frac{m_a m_b}{m_c}\mom{c}{k}\spin{c}{\ell}{m}\nonumber\\
&&\times\partial_i^{(a)} \partial_j^{(b)} \partial_m^{(c)}\biggl[\biggl(\delta_{k(i}\delta_{j)\ell}-\frac{1}{2}\delta_{ij}\delta_{k\ell}\biggr)
\ln s_{abc}\nonumber\\
&&+\frac{1}{2}\biggl(\frac{1}{2}\delta_{k\ell}\partial_i^{(c)} \partial_j^{(c)} - \delta_{\ell(i}\partial_{j)}^{(c)}\partial_k^{(c)}\biggr)\nonumber\\
&&\biggl\{\frac{1}{18}(3 (r_{ab} r_{ac} + r_{ab} r_{bc}\nonumber\\
&&\quad -  r_{ac} r_{bc})-r_{ac}^2 - r_{bc}^2  + r_{ab}^2) \nonumber\\
&&+ \frac{1}{6} (r_{ac}^2 + r_{bc}^2 - r_{ab}^2)\ln s_{abc}\biggr\}\biggr]\,,\\
 H^{\text{matter,SS}}_{\phi(2)\pi(3)^2} & = & 2\gravthree^2 \sum_{a}\sum_{b\ne a}\sum_{c\ne a,b} m_c\nonumber\\
&&\times\biggl[ \spin{a}{k}{i}\spin{b}{\ell}{i}\partial_{k}^{(a)}\partial_{j}^{(a)}\partial_{\ell}^{(b)}\partial_{j}^{(b)}\\
&&+\spin{a}{k}{i}\spin{b}{\ell}{j}\partial_{k}^{(a)}\partial_{j}^{(a)}\partial_{\ell}^{(b)}\partial_{i}^{(b)}\biggr]\ln s_{abc}\,.\nonumber
\end{eqnarray}
Because of the two-body and one-body contributions in a three-body sum,
one has to decompose the sum into a purely one-body part, a two-body part and a three-body part.
The first two parts can be calculated by dimensional regularization procedures via the Riesz formula
mentioned in \cite{Damour:Jaranowski:Schafer:2001} 
and the three-body integrals were calculated by applying the formulas given above. The delta-type one- and two-body contributions were calculated by inserting Riesz kernel regulators and applying the method of dimensional regularization. The sources and the constraint decomposition in $\dim$ dimensions will be given in a forthcoming 
publication.

To eliminate the scalar products of $\vnxa{ab}$, $\vnxa{bc}$ and $\vnxa{ac}$ (here $\vnxa{ab} = (\vx{a}-\vx{b})/r_{ab}$) one can make use of the following identities
\begin{eqnarray}
 (\vnxa{ac}\vnxa{bc}) & = & \frac{r_{ac}^2 + r_{bc}^2 - r_{ab}^2}{2 r_{ac} r_{bc}}\,,\\
 (\vnxa{ab}\vnxa{bc}) & = & -\frac{r_{ab}^2 + r_{bc}^2 - r_{ac}^2}{2 r_{ab} r_{bc}}\,,\\
 (\vnxa{ab}\vnxa{ac}) & = & \frac{r_{ab}^2 + r_{ac}^2 - r_{bc}^2}{2 r_{ab} r_{ac}}\,,
\end{eqnarray}
additional to $\vnxa{ab}^2 = 1$.
Note the minus sign in front of the second expression, which comes from relabeling the first identity and changing the direction of the second factor in the scalar product.

Furthermore one can remove one of the appearing unit vectors due to the fact that they are not linearly independent, namely
\begin{eqnarray}
 \vnxa{ab} & = & \frac{r_{ac}}{r_{ab}}\vnxa{ac}-\frac{r_{bc}}{r_{ab}}\vnxa{bc}\,,\\
\vnxa{ac} & = & \frac{r_{ab}}{r_{ac}}\vnxa{ab}+\frac{r_{bc}}{r_{ac}}\vnxa{bc}\,,\\
\vnxa{bc} & = & -\frac{r_{ab}}{r_{bc}}\vnxa{ab}+\frac{r_{ac}}{r_{bc}}\vnxa{ac}\,.
\end{eqnarray}

%% file: 3SpinPaperRes.tex
\section{Results}\label{sec:Results}
In this section we make use of {\scshape xTensor} \cite{MartinGarcia:2002}, a free package
for {\scshape Mathematica} \cite{Wolfram:2003}, especially of its fast index
canonicalizer based on the package {\scshape xPerm} \cite{MartinGarcia:2008},
and some of our own code for evaluating integrals and derivatives.

The Hamiltonians consist of several parts, which will be provided below, namely
\begin{subequations}
\begin{eqnarray}
 H^{\text{NLO}}_{\text{SO}} & = & H^{[2],\gravthree}_{\text{NLO,SO}} + H^{[2],\gravthree^2}_{\text{NLO,SO}} + H^{[3],\gravthree^2}_{\text{NLO,SO}}\,,\\
 H^{\text{NLO}}_{\text{SS}} & = & H^{[2],\gravthree}_{\text{NLO,SS}} + H^{[2],\gravthree^2}_{\text{NLO,SS}} + H^{[3],\gravthree^2}_{\text{NLO,SS}}\,,
\end{eqnarray}
\end{subequations}
where the numbers in square brackets denotes the number of compact objects involved in the interaction.
\subsection{Spin-orbit Hamiltonian}
\subsubsection{two-body interaction part}
The two-body interaction parts of the spin-orbit Hamiltonian linear in $\gravthree$ and quadratic in $\gravthree$ are given by
\begin{widetext}
\begin{eqnarray}
 H_{\text{NLO,SO}}^{[2],\,\gravthree} 
&=&  \sum_a \sum_{b\ne a} \frac{\gravthree}{r_{ab}^2}\biggl[\biggl(\frac{3}{4}\frac{\vmom{b}^2}{m_a m_b}-\frac{3}{2}\frac{(\vnxa{ab}\vmom{b})^2}{m_a m_b} -\frac{5}{8} \frac{m_b \vmom{a}^2}{m_a^3}-\frac{3}{4} \frac{(\vmom{a}\vmom{b})}{m_a^2} - \frac{3}{4} \frac{(\vnxa{ab}\vmom{a})(\vnxa{ab}\vmom{b})}{m_a^2}\biggr)\sppr{\vnxa{ab}}{\vmom{a}}{\vspin{a}}\nonumber\\
&&+\biggl(\frac{(\vmom{a}\vmom{b})}{m_a m_b} + 3 \frac{(\vnxa{ab}\vmom{a})(\vnxa{ab}\vmom{b})} {m_a m_b}\biggr)\sppr{\vnxa{ab}}{\vmom{b}}{\vspin{a}}
+\biggl(\frac{3}{4}\frac{(\vnxa{ab}\vmom{a})}{m_a^2}-2\frac{(\vnxa{ab}\vmom{b})}{m_a m_b}\biggr)\sppr{\vmom{a}}{\vmom{b}}{\vspin{a}}
\biggr]\,,\\
 H_{\text{NLO,SO}}^{[2],\,\gravthree^2} & = &  \sum_a \sum_{b\ne a} \frac{\gravthree^2}{r_{ab}^3}\biggl[-\left(\frac{11}{2}m_b + 5\frac{m_b^2}{m_a}\right)\sppr{\vnxa{ab}}{\vmom{a}}{\vspin{a}}+\left(6 m_a + \frac{15}{2}m_b\right)\sppr{\vnxa{ab}}{\vmom{b}}{\vspin{a}}\biggr]\,,
\end{eqnarray}
and consist of two-body interaction parts of the field integral  $H_{\phi(2)\pi(3)^2}$ and a part of the delta-type integrals. Comparison with $1/r_{12}^2$ and $1/r_{12}^3$ terms of the next-to-leading order spin-orbit Hamiltonian in \cite{Damour:Jaranowski:Schafer:2008:1,Steinhoff:Schafer:Hergt:2008} leads to full agreement.
\subsubsection{three-body interaction part}
The three-body interaction Hamiltonian is always at $\gravthree^2$ level and consists of three different parts:
\begin{eqnarray}
 H^{[3],\gravthree^2}_{\text{NLO,SO}} & = & H_{\delta-\text{type}}^{\text{matter,SO}}+H^{\text{matter,SO}}_{\phi(2)\pi(3)^2}+H^{\text{TT-part,SO}}_{\partial \phi(2)^2\htt{}{(4)}}\,.
\end{eqnarray}
The delta-type part (which results from the source parts of the integrand) is given by
\begin{eqnarray}
H_{\delta-\text{type}}^{\text{matter,SO}}
&=&-\sum_{a}\sum_{b\ne a}\sum_{c\ne a,b} \frac{\gravthree^2}{r_{ab}^2}\left(\frac{5}{r_{ac}}+\frac{1}{r_{bc}}\right)\frac{m_b m_c}{m_a}\sppr{\vnxa{ab}}{\vmom{a}}{\vspin{a}}\,.
\label{eq:3partdelta}
\end{eqnarray}
The pure field part is given by 
\begin{eqnarray}
 H^{\text{matter,SO}}_{\phi(2)\pi(3)^2} &=& \sum_{a} \sum_{b\ne a}\sum_{c\ne a,b} \gravthree^2 m_a\biggl[
\frac{1}{s_{abc}} \biggl\{
	\biggl( -\frac{8}{r_{bc}^2}
		+ \frac{8}{r_{ab}r_{ac}}
		- \frac{4}{r_{ab}r_{bc}}
		- \frac{4}{r_{ac}r_{bc}}
		- \frac{4 r_{ab}}{r_{ac} r_{bc}^2}
		- \frac{4 r_{ac}}{r_{ab} r_{bc}^2}
	\biggr) \sppr{\vnxa{bc}}{\vmom{b}}{\vspin{c}} \nla
	+ \biggl( \frac{3}{r_{ab}r_{ac}}
		- \frac{6}{r_{ab}r_{bc}}
		- \frac{3}{r_{ac}r_{bc}}
		- \frac{3 r_{ab}}{r_{ac}^2 r_{bc}}
		+ \frac{3 r_{bc}}{r_{ab} r_{ac}^2}
	\biggr) \sppr{\vnxa{ac}}{\vmom{b}}{\vspin{c}}
\biggr\}
+ \frac{\sppr{\vnxa{ac}}{\vnxa{bc}}{\vspin{c}}}{s_{abc}^2} \biggl\{
	\frac{16 (\vnxa{ac}\vmom{b})}{r_{ab}} \nla
	+ \biggl( \frac{4}{r_{ab}}
		- \frac{2 r_{bc}}{r_{ab} r_{ac}}
		+ \frac{2 r_{ac}}{r_{ab}^2}
		- \frac{2 r_{bc}^2}{r_{ab}^2 r_{ac}}
		+ \frac{8 r_{ab}}{r_{ac}^2}
		+ \frac{7 r_{bc}}{r_{ac}^2}
		- \frac{2 r_{bc}^2}{r_{ab} r_{ac}^2}
		+ \frac{r_{bc}}{r_{ab}^2}
		- \frac{r_{bc}^3}{r_{ab}^2 r_{ac}^2}
	\biggr) (\vnxa{ab}\vmom{b}) \nla
	+\biggl( \frac{12}{r_{ab}}
		- \frac{2}{r_{ac}}
		+ \frac{5}{r_{bc}}
		- \frac{2 r_{ab}}{r_{ac}^2}
		+ \frac{7 r_{bc}}{r_{ac}^2}
		- \frac{2 r_{ab}}{r_{ac}r_{bc}}
		+ \frac{6 r_{ac}}{r_{ab} r_{bc}}
		- \frac{r_{ab}^2}{r_{ac}^2 r_{bc}}
		+ \frac{8 r_{bc}^2}{r_{ab} r_{ac}^2}
	\biggr) (\vnxa{bc}\vmom{b})
\biggr\}
 \biggr] \,.
\end{eqnarray}
The TT field part of this Hamiltonian is given by
\begin{eqnarray}
 H^{\text{TT-part,SO}}_{\partial \phi(2)^2\htt{}{(4)}} & = & \sum_a \sum_{b\ne a} \sum_{c \ne a,b} \gravthree^2 \frac{m_a m_b}{m_c} \biggl[\frac{1}{s_{abc}^2}\biggl\{-\frac{1}{r_{ac}}-\frac{1}{r_{ab}}\left(2+\frac{1}{2}\frac{r_{bc}}{r_{ac}}\right)
+\frac{1}{r_{ab}^2}\left(-4r_{ac}+r_{bc}+\frac{r_{bc}^2}{r_{ac}}\right)\nonumber\\
&&+\frac{1}{r_{ab}^3}\left(-2r_{ac}^2+r_{bc}^2-\frac{3}{2}r_{ac}r_{bc}+\frac{1}{2}\frac{r_{bc}^3}{r_{ac}}\right)\biggr\}
(\vnxa{ac}\vmom{c})\sppr{\vnxa{ac}}{\vnxa{bc}}{\vspin{c}}\nonumber\\
&&+\frac{1}{s_{abc}}\biggl\{\frac{1}{8 r_{ac}^2}-\frac{1}{4 r_{ac} r_{bc}}-\frac{r_{ab}}{4 r_{ac}^2 r_{bc}}
+\frac{1}{r_{ab}}\left(\frac{3}{8 r_{ac}} - \frac{1}{r_{bc}}+\frac{3}{8}\frac{r_{bc}}{r_{ac}^2}\right)\nonumber\\
&&+\frac{1}{r_{ab}^2}\left(-\frac{5}{8}+\frac{3}{4}\frac{r_{ac}}{r_{bc}}-\frac{1}{8}\frac{r_{bc}^2}{r_{ac}^2}\right)
+\frac{1}{r_{ab}^3}\left(\frac{1}{8} r_{ac} - \frac{5}{8} r_{bc} + \frac{3}{4}\frac{r_{ac}^2}{r_{bc}}
-\frac{1}{8}\frac{r_{bc}^2}{r_{ac}}-\frac{1}{8}\frac{r_{bc}^3}{r_{ac}^2}\right)\biggr\}\
\sppr{\vnxa{ac}}{\vmom{c}}{\vspin{c}}\nonumber\\
&&+(a\leftrightarrow b)\biggr]\,,
\end{eqnarray}
where $(a\leftrightarrow b)$ denotes an exchange of object labels $a$ and $b$ in all of the preceding terms.

\subsection{Spin(a)-Spin(b) Hamiltonian}
\subsubsection{two-body interaction part}
The whole spin(a)-spin(b) Hamiltonian linear in $\gravthree$ (which is a sum of the delta-type part, the TT-part, and the vector potential part) is given by
\begin{eqnarray}
 H^{[2],\gravthree}_{\text{NLO,SS}} & = & \sum_a \sum_{b\ne a}\frac{\gravthree}{r_{ab}^3}\biggl[\frac{1}{4 m_a m_b}\biggl(6\sppr{\vnxa{ab}}{\vmom{b}}{\vspin{a}}\sppr{\vnxa{ab}}{\vmom{a}}{\vspin{b}}+\frac{3}{2}\sppr{\vnxa{ab}}{\vmom{a}}{\vspin{a}}\sppr{\vnxa{ab}}{\vmom{b}}{\vspin{b}}\nonumber\\
&&\quad-15(\vnxa{ab}\vmom{a})(\vnxa{ab}\vmom{b})(\vnxa{ab}\vspin{a})(\vnxa{ab}\vspin{b})-3(\vmom{a}\vmom{b})(\vnxa{ab}\vspin{a})(\vnxa{ab}\vspin{b})+3(\vnxa{ab}\vmom{b})(\vmom{a}\vspin{a})(\vnxa{ab}\vspin{b})\nonumber\\
&&\quad+3(\vnxa{ab}\vmom{a})(\vmom{b}\vspin{a})(\vnxa{ab}\vspin{b})+3(\vnxa{ab}\vmom{b})(\vnxa{ab}\vspin{a})(\vmom{a}\vspin{b})+3(\vnxa{ab}\vmom{a})(\vnxa{ab}\vspin{a})(\vmom{b}\vspin{b})\nonumber\\
&&\quad-\frac{1}{2}(\vmom{b}\vspin{a})(\vmom{a}\vspin{b})+(\vmom{a}\vspin{a})(\vmom{b}\vspin{b})-3(\vnxa{ab}\vmom{a})(\vnxa{ab}\vmom{b})(\vspin{a}\vspin{b})+\frac{1}{2}(\vmom{a}\vmom{b})(\vspin{a}\vspin{b})\biggr)\nonumber\\
&&+\frac{3}{2 m_a^2}\biggl(-\sppr{\vnxa{ab}}{\vmom{a}}{\vspin{a}}\sppr{\vnxa{ab}}{\vmom{a}}{\vspin{b}}-(\vnxa{ab}\vmom{a})(\vnxa{ab}\vspin{a})(\vmom{a}\vspin{b})+(\vnxa{ab}\vmom{a})^2(\vspin{a}\vspin{b})\biggr)\biggr]\,.
\end{eqnarray}
The $\gravthree^2$ two-body interaction Hamiltonian is given by
\begin{eqnarray}
 H^{[2],\gravthree^2}_{\text{NLO,SS}} & = & \sum_a \sum_{b\ne a}\frac{6\gravthree^2 m_a }{r_{ab}^4} \left[(\vspin{a}\vspin{b})-2(\vnxa{ab}\vspin{a})(\vnxa{ab}\vspin{b})\right]\,.
\end{eqnarray}
We neglected appearing $\vspin{a}^2$ terms due to consistency reasons (the stress-energy tensor does not contain $\vspin{a}^2$ expressions). Note that these two parts of the Hamiltonian are also in perfect agreement with \cite{Steinhoff:Hergt:Schafer:2008:2,Steinhoff:Schafer:Hergt:2008,Levi:2008}.

\subsubsection{three-body interaction part}
The three-body interaction part of this Hamiltonian is given by $H^{[3],\gravthree^2}_{\text{NLO,SS}}=H^{\text{matter,SS}}_{\phi(2)\pi(3)^2}$, namely
\begin{eqnarray}
H^{\text{matter,SS}}_{\phi(2)\pi(3)^2} & = & \sum_{a} \sum_{b\ne a}\sum_{c\ne a,b}\frac{\gravthree^2 m_c}{s_{abc}^2}\biggl[
	\sppr{\vnxa{ac}}{\vnxa{bc}}{\vspin{a}}\sppr{\vnxa{ac}}{\vnxa{bc}}{\vspin{b}}
	\biggl\{
		\frac{1}{r_{ac}r_{bc}}
		+\frac{4}{r_{ab} r_{ac}}
		+\frac{4}{r_{ab}^2}
		\left(
			\frac{1}{2}
			+\frac{r_{ac}}{r_{bc}}
		\right)\nonumber\\
		&&+\frac{2}{r_{ab}^3}
		\left(
			2r_{ac}
			+\frac{r_{ac}^2}{r_{bc}}
		\right)
	\biggr\}+(\vnxa{ac}\vspin{a})(\vnxa{ac}\vspin{b})\biggl\{
		\frac{2}{r_{ac}r_{bc}}
		+\frac{1}{r_{ab}}
		\left(
			\frac{1}{r_{ac}}
			+\frac{4}{r_{bc}}
		\right)
		-\frac{2}{r_{ab}^2}
		\left(
			1
			-\frac{2 r_{ac}}{r_{bc}}
			+\frac{r_{bc}}{r_{ac}}
		\right)\nonumber\\
		&&-\frac{1}{r_{ab}^3}
		\left(
			2 r_{ac}
			+2 r_{bc}
			+\frac{r_{ac}^2}{r_{bc}}
			+\frac{r_{bc}^2}{r_{ac}}
		\right)
		-\frac{6}{r_{ab}^4}
		\left(
			2 r_{ac}^2
			+r_{ac} r_{bc}
			+\frac{r_{ac}^3}{r_{bc}}
		\right)
		-\frac{3}{r_{ab}^5}
		\left(
			3 r_{ac}^3
			+3 r_{ac}^2 r_{bc}
			+r_{ac} r_{bc}^2
			+\frac{r_{ac}^4}{r_{bc}}
		\right)
	\biggr\}\nonumber\\
	&&+(\vnxa{bc}\vspin{a})(\vnxa{ac}\vspin{b})\biggl\{
		-\frac{1}{r_{ac}^2}
		-\frac{1}{r_{ac} r_{bc}}
		-\frac{2 r_{ab}}{r_{ac}^2 r_{bc}}
		-\frac{r_{ab}^2}{2 r_{ac}^2 r_{bc}^2}
		+\frac{r_{ac}}{r_{ab} r_{bc}^2}
		+\frac{2}{r_{ab}^2}
		\left(
			-1
			+\frac{r_{ac}}{r_{bc}}
			+\frac{r_{ac}^2}{r_{bc}^2}
		\right)\nonumber\\
		&&+\frac{1}{r_{ab}^3}
		\left(
			2 r_{ac}
			+\frac{2 r_{ac}^2}{r_{bc}}
			+\frac{r_{ac}^3}{r_{bc}^2}
		\right)
		+\frac{3(r_{ac}+r_{bc})^2}{r_{ab}^4}
		+\frac{3(r_{ac}+r_{bc})^3}{2 r_{ab}^5}
	\biggr\}
	+(\vnxa{ac}\vspin{a})(\vnxa{bc}\vspin{b})\biggl\{
		-\frac{2}{r_{ab}^2}
		+\frac{r_{ac}}{r_{ab}^3}\nonumber\\
		&&+\frac{3(r_{ac}+r_{bc})^2}{r_{ab}^4}
		+\frac{3(r_{ac}+r_{bc})^3}{2 r_{ab}^5}
	\biggr\}
	+(\vspin{a}\vspin{b})\biggl\{
		\frac{2}{r_{ac}^2}
		-\frac{3}{2 r_{ac} r_{bc}}
		+\frac{3}{2}\frac{r_{ac}}{r_{bc}^3}
		+r_{ab}
		\left(
			\frac{3}{2 r_{ac}^3}
			+\frac{1}{r_{ac}^2 r_{bc}}
		\right)
		-\frac{r_{ab}^2}{2 r_{ac}^2 r_{bc}^2}\nonumber\\
		&&-\frac{r_{ab}^3}{r_{ac}^3 r_{bc}^2}
		-\frac{r_{ab}^4}{4 r_{ac}^3 r_{bc}^3}
		+\frac{1}{r_{ab}}
		\left(
			-\frac{2}{r_{ac}}
			+\frac{r_{ac}}{r_{bc}^2}
		\right)
		+\frac{1}{r_{ab}^2}
		\left(
			3
			+\frac{3 r_{ac}}{r_{bc}}
			-\frac{r_{ac}^2}{r_{bc}^2}
			-\frac{r_{ac}^3}{r_{bc}^3}
		\right)\nonumber\\
		&&+\frac{1}{r_{ab}^3}
		\left(
			\frac{9}{2}r_{ac}
			+\frac{r_{ac}^2}{r_{bc}}
			-\frac{r_{ac}^3}{r_{bc}^2}
			-\frac{r_{ac}^4}{2 r_{bc}^3}
		\right)
	\biggr\}
	+(a\leftrightarrow b)\biggr]\,.
\end{eqnarray}

\subsection{Approximate Poincar\'e algebra}\label{subsec:Poincare}
As a check of the Hamiltonians given in the previous section we look
at the global Poincar\'e algebra, see, e.g., \cite{Damour:Jaranowski:Schafer:2008:1}. For this we need the
center of mass vector
\begin{equation}
\vct{G} = - \frac{1}{16\pi\gravthree} \int \text{d}^3x \, \vct{x}\, \Delta
	\phi \,. 
\end{equation}
Since the center of mass vector integrals are given by the Hamilton constraint equations which are one order below the appropriate integrals for the Hamiltonians, there are no explicit three-body parts appearing there. For the readers convenience we provide them here as well. (Notice the abuse of vocabulary, in fact $\vct{G}/H$ is the center of mass, but we refer to $\vct{G}$ as center of mass vector.) The Newtonian center of mass vector can be calculated trivially and is given by
\begin{eqnarray}
 \vct{G}^{\text{N}} & = & \sum_a m_a \vx{a}\,.
\end{eqnarray}
The 1PN point-mass center of mass vector is given by
\begin{eqnarray}
 \vct{G}^{\text{1PN}}_{\text{PP}} & = & \sum_a \frac{\vmom{a}^2}{2 m_a} \vx{a} - \frac{1}{2} \sum_a \sum_{b\ne a} \frac{\gravthree m_a m_b}{r_{ab}} \vx{a}\,,
\end{eqnarray}
see, e.g., \cite{Landau:Lifshitz:Vol2:2,Damour:Jaranowski:Schafer:2000,*Damour:Jaranowski:Schafer:2000:err}.
The leading order spin-orbit center of mass vector is given by
\begin{eqnarray}
\vct{G}^{\text{LO}}_{\text{SO}} & = & \sum_a \frac{1}{2 m_a} \crpr{\vmom{a}}{\vspin{a}} \,.
\end{eqnarray}
There exists no leading order spin(a)-spin(b) center of mass vector. The next-to-leading order spin-orbit center of mass vector is given by
\begin{eqnarray}
 \vct{G}^{\text{NLO}}_{\text{SO}} & = & -\sum_a \frac{\vmom{a}^2}{8 m_a^3}\crpr{\vmom{a}}{\vspin{a}}+\sum_a \sum_{b\ne a} \frac{\gravthree}{r_{ab}}\frac{m_b}{4 m_a}\biggl[-5\crpr{\vmom{a}}{\vspin{a}}
+\sppr{\vmom{a}}{\vspin{a}}{\vnxa{ab}}\frac{5\vx{a}+\vx{b}}{r_{ab}}\biggr] \nla
+\sum_a \sum_{b\ne a}\frac{\gravthree}{r_{ab}}\biggl[\frac{3}{2}\crpr{\vmom{b}}{\vspin{a}}-\frac{1}{2}\crpr{\vnxa{ab}}{\vspin{a}}(\vmom{b}\vnxa{ab})-\sppr{\vmom{b}}{\vspin{a}}{\vnxa{ab}}\frac{\vx{a}+\vx{b}}{r_{ab}}\biggr]\,,
\end{eqnarray}
and the next-to-leading order spin(a)-spin(b) center of mass vector is given by
\begin{eqnarray}
 \vct{G}^{\text{NLO}}_{\text{SS}} & = & \frac{\gravthree}{2}\sum_a \sum_{b\ne a} \biggl[(\vspin{b}\vnxa{ab})\frac{\vspin{a}}{r_{ab}^2} + (3(\vspin{a}\vnxa{ab})(\vspin{b}\vnxa{ab})-(\vspin{a}\vspin{b}))\frac{\vx{a}}{r_{ab}^3}\biggr]\,,
\end{eqnarray}
see, e.g., \cite{Damour:Jaranowski:Schafer:2008:1,Steinhoff:Hergt:Schafer:2008:2,Steinhoff:Schafer:Hergt:2008}
(notice that there is a misprint in $\vct{G}^{\text{NLO}}_{\text{SO}}$ in the published version of  \cite{Steinhoff:Schafer:Hergt:2008}). For comparison of the $\gravthree^0$ parts (up to linear order in spin) of the center of mass vectors and Hamiltonians one can use the center of mass vector and the Hamiltonian which can be calculated by integrating the source $\src{}$ directly and setting $\gamma_{ij} = \delta_{ij}$, which results in
\begin{eqnarray}
 G^i_{\text{SRT}} & = & \sum_a \biggl[\sqrt{m_a^2 + \vmom{a}^2}\xa{a}{i} + \frac{\mom{a}{\ell}\spin{a}{i}{\ell}}{m_a + \sqrt{m_a^2 + \vmom{a}^2}}\biggr]\,,\\
 H_{\text{SRT}} & = & \sum_a \sqrt{m_a^2 + \vmom{a}^2}\,.
\end{eqnarray}
From the references given above and, e.g., \cite{Einstein:Infeld:Hoffmann:1938,Landau:Lifshitz:Vol2:2,Lousto:Nakano:2008}
one can also get the Hamiltonians needed, which are given by
\begin{eqnarray}
 H^{\text{N}} & = & \sum_a \frac{\vmom{a}^2}{2 m_a} -\frac{1}{2} \sum_a \sum_{b\ne a} \frac{\gravthree m_a m_b}{r_{ab}}\,,\\
 H^{\text{1PN}}_{\text{PP}} & = & -\sum_a \frac{(\vmom{a}^2)^2}{8 m_a^3} + \sum_a \sum_{b\ne a} \frac{\gravthree}{r_{ab}}\biggl[-\frac{3 m_b}{2 m_a} \vmom{a}^2 + \frac{1}{4}\left(7(\vmom{a}\vmom{b}) + (\vnxa{ab}\vmom{a})(\vnxa{ab}\vmom{b})\right)\biggr] + \sum_a \sum_{b\ne a} \frac{\gravthree^2 m_a^2 m_b}{2 r_{ab}^2}\nonumber\\
&&+ \sum_a \sum_{b\ne a} \sum_{c\ne a,b} \frac{\gravthree^2 m_a m_b m_c}{2 r_{ab} r_{ac}}\,,\\
 H^{\text{LO}}_{\text{SO}} & = & \sum_a \sum_{b\ne a} \frac{\gravthree}{r_{ab}^2} \biggl[\frac{3 m_b}{2 m_a} \sppr{\vnxa{ab}}{\vmom{a}}{\vspin{a}}-2\sppr{\vnxa{ab}}{\vmom{b}}{\vspin{a}}\biggr]\,,\\
 H^{\text{LO}}_{\text{SS}} & = & \sum_a \sum_{b\ne a} \frac{\gravthree}{2 r_{ab}^3}\biggl[3(\vnxa{ab}\vspin{a})(\vnxa{ab}\vspin{b})-(\vspin{a}\vspin{b})\biggr]\,.
\end{eqnarray}
\end{widetext}
It is now straightforward, though rather lengthy, to check that the global Poincar\'e algebra is fulfilled. Note that for checking spin-orbit parts of the $\{\vct{G},H\}$ part of the Poincar\'{e} algebra for example one also has to include  $\{\vct{G}_{\text{SO}},H_{\text{SO}}\}$ due to the spin Poisson brackets.

The Hamiltonians given above are necessary to check the Poincar\'e algebra relations 
involving the derived Hamiltonians.
They are not sufficient to simulate the full post-Newtonian dynamics at 2.5PN. Additionally to
the derived next-to-leading order spin-orbit Hamiltonian and
the Hamiltonians mentioned above, one needs the three-body 2PN point-mass Hamiltonian \cite[Eq. (5)]{Schafer:1987}
and \cite[Eq. (A.1)]{Lousto:Nakano:2008}, the leading order spin(a)-spin(a) Hamiltonian
\begin{eqnarray}
 H^{\text{LO}}_{\text{S}^2} & = & \sum_a \sum_{b\ne a} \frac{G}{2 r_{ab}^3 }\frac{m_b}{m_a}C_{Q\,a}\left[3(\vspin{a}\vnxa{ab})^2-\vspin{a}^2\right]\,,
\end{eqnarray}
given in, e.g., \cite[Eq. (13)]{Hergt:Steinhoff:Schafer:2010:1} (the constant $C_{Q\,a}$ parametrizing the quadrupole deformation due to spin for the $a$th object, with $C_{Q\,a}=1$ for a black hole), and the radiative 2.5PN point-mass Hamiltonian provided in \cite[Eq. (41)]{Jaranowski:Schafer:1997}.

\subsection{Another derivation of the Hamiltonians}
Because of the momentum independence of the three-body part of the spin(a)-spin(b) Hamiltonian,
one cannot check this part using the Poincar\'{e} algebra. Therefore we rederived the Hamiltonians
$H^{\text{NLO}}_{\text{SO}} = \sum_a \vct{\Omega}_{a(4)}\vspin{a}$ and $H^{\text{NLO}}_{\text{SS}}$
using the formalism given in \cite{Damour:Jaranowski:Schafer:2008:1}
via the precession frequency $\Omega_{a(4)}^{i}$, Eq.\ (4.10) in \cite{Damour:Jaranowski:Schafer:2008:1},
and compared this with our result given above. (Note that for the spin(a)-spin(b)
Hamiltonian it is in principle not necessary to derive all $\vct{\Omega}_a$.
One only needs the term containing one of the spins. So one has to multiply a factor $1/2$ when adding up
all parts of the Hamiltonian namely $H^{\text{NLO}}_{\text{SS}} = \tfrac{1}{2}\sum_a \vct{\Omega}_{(4)a} \vspin{a}$
to avoid overcounting.) Both results for spin(a)-spin(b) and spin-orbit are \emph{identical}
with our previous results. It is explained in the appendix why they should not even differ by
a canonical transformation. The field variables necessary for calculating the spin
precession frequency were taken from \cite{Schafer:1985,Jaranowski:Schafer:1998,Steinhoff:Schafer:Hergt:2008}.%
\footnote{Note there is a misprint in Eq. (5.6b) in \cite{Schafer:1985} in the term
$2 \tfrac{1}{r_a r_{ab}} \{\dots\}$. This term should be $2 \tfrac{1}{r_b r_{ab}}
\{\dots\}$ and in the last term before $\tilde{I}^{(5)}_{,i}$ the free index of
$\vnxa{b}$ should be $i$ instead of $j$.}

Notice that the performed rederivation of the Hamiltonians via the spin precession frequency provides
a very strong check because it needs expressions for lapse and shift,
which are eliminated in the formalism we used before.

%% file: 3SpinPaperCon.tex
\section{Conclusions and Outlook\label{sec:Conclusions}}
We have derived the post-Newtonian next-to-leading order conservative spin-orbit and
spin(a)-spin(b) gravitational interaction Hamiltonians for arbitrary many
compact objects. The spin-orbit Hamiltonian completes the knowledge
of Hamiltonians up to and including 2.5PN for three compact rapidly rotating
objects. The Hamiltonians were checked with the help of the
Poincar\'e algebra and rederived with the independent method from
\cite{Damour:Jaranowski:Schafer:2008:1}.


A possible astro-physical application of our computation should be
the exploration of Kozai resonance in hierarchical triples containing
spinning compact objects in a fully post-Newtonian accurate manner.
Recall that in hierarchical triples experiencing Kozai resonance
the orbital eccentricity of the inner binary secularly evolves, mainly
due to the tidal torquing between the inclined inner and outer orbits
\cite{Kozai:1962}. And, the general relativistic periastron advance
of the inner binary can interfere with, and in principle terminate,
the evolution of its eccentricity \cite{Ford:Kozinsky:Rasio:2000,*Ford:Kozinsky:Rasio:2000:err}.
Therefore, the present computation should be useful in extending the detailed analysis presented
in \cite{Ford:Kozinsky:Rasio:2000,*Ford:Kozinsky:Rasio:2000:err}.
Interestingly, we note that Kozai resonance, as discussed in
Ref.~\cite{Ford:Kozinsky:Rasio:2000,*Ford:Kozinsky:Rasio:2000:err},
is also proposed as a scenario to merge massive binary black holes
resulting from galaxy mergers \cite{Blaes:Lee:Socrates:2002}.

To best of our knowledge not even the leading order spin Hamiltonians were applied
in this context. Though the next-to-leading order effects derived here
are considerably weaker within the validity of the post-Newtonian approximation, they
can still be important. For the three-body case many configurations are
potentially chaotic and weak interaction terms can have big effects.
Further the leading order spin Hamiltonians only consist of two-body
interactions, i.e., the objects interact pairwise with each other as
in Newtonian gravity. At the next-to-leading order the most complicated
parts of the Hamiltonians are three-body interactions and thus provide not just
a refinement of the leading order dynamics. At the next-to-leading
order the complexity of Einstein's theory of gravitation becomes apparent.
Finally, the size of next-to-leading order effects provides a handle on
the accuracy of the leading order.

Additionally, it is interesting to see whether one can extend known three-body solutions
without spin, see, e.g., \cite{Moore:1993,Imai:Chiba:Asada:2007,Lousto:Nakano:2008},
to the three-body problem with spin at certain order.
In the literature, there exist parametrizations for the binary case at leading order spin-orbit
\cite{Konigsdorffer:Gopakumar:2005,Tessmer:2009}.
For special configurations like spins aligned to orbital angular momentum, a parametrization
for three bodies including next-to-leading order spin-orbit interaction seems to be possible,
see \cite{Tessmer:Hartung:Schafer:2010} for the binary case.

To foster application of the derived Hamiltonians we provide them for three compact objects as
Mathematica source files \cite{sourcefiles}.

%% file: 3SpinPaperApp.tex
\appendix{
\section{Relation to the spin variable used by Damour, Jaranowski, and Sch\"afer}
In the past \cite{Steinhoff:Schafer:Hergt:2008} the method in \cite{Damour:Jaranowski:Schafer:2008:1} was already used as a check,
but it was not clear why the Hamiltonians were in perfect agreement (and not differing by a canonical transformation).
To explain this issue we compare the canonical spin $\vspin{a}^{\text{DJS}}$ used
in \cite{Damour:Jaranowski:Schafer:2008:1} and the canonical
spin $\vspin{a}$ used in the present paper and in \cite{Steinhoff:Schafer:Hergt:2008}.

The comparison was done in the following way. We constructed the matrix $G^{ij}$ due to Eq. (2.7) in \cite{Damour:Jaranowski:Schafer:2008:1}.
From that we calculated the symmetric matrix square root $H^{ij}$ which
relates $\vspin{a}^{\text{DJS}}$ to the spatial components of the covariant spin 4-vector $S_{a\,\mu}$ 
(which fulfills the covariant spin supplementary condition $S_{a\,\mu} u^\mu_a = 0$).\footnote{
Note that in \cite{Damour:Jaranowski:Schafer:2008:1} the canonical spin is $\vct{S}_a$ (denoted
by $\vspin{a}^{\text{DJS}}$ here) and the covariant spin $\tilde{\vct{S}}_a$, whereas in
\cite{Steinhoff:Schafer:Hergt:2008} the canonical spin is $\vspin{a}$ and the covariant spin $\vct{S}_a$. 
We use the convention of \cite{Steinhoff:Schafer:Hergt:2008} here.}
Notice that it is enough to compare the definitions of the canonical spin variables,
as the formalism in \cite{Damour:Jaranowski:Schafer:2008:1} is based on the
spin equation of motion, in which corrections to the canonical position
and momentum are of higher order in spin and can be neglected.

Now we split up the equation which relates $S_{a\,\mu}$ to the spin tensor
$S^{\mu\nu}_{a}$ given by
\begin{eqnarray}
 S_{a\,\mu} &=& \frac{1}{2} \sqrt{-g_{(4)}}\epsilon_{\mu\nu\alpha\beta} u_a^\nu S^{\alpha\beta}_{a}\,,
\end{eqnarray}
in a (3+1) manner. This gives
\begin{eqnarray}
 S_{a\,i} &=& \frac{\sqrt{\gamma}}{m_a} \epsilon_{ijk} \gamma^{jm}\gamma^{kn}
\left(\mom{a}{m} (nS_a)_{n} - \frac{1}{2} nP_a S_{a\,mn}\right)\,,\nonumber\\\label{eq:31stilde}
\end{eqnarray}
using the (3+1) decomposition of $n^\mu$, $\epsilon^{0123} = 1$ such that $\epsilon_{0123} = -1$ and so
$\epsilon_{0ijk} = -\epsilon_{ijk}$, and $u^\nu_a = P_a^\nu/m_a$. After the (3+1) split we insert $(nS_a)_i = -\mom{a}{k}\gamma^{kj}\hat{S}_{a\,ji}/m_a$, $nP_a = -\sqrt{m_a^2 + \gamma^{ij}\mom{a}{i}\mom{a}{j}}$ and the
transformation from the covariant spin to the Newton-Wigner spin, namely
\begin{eqnarray}
 S_{a\,ij} &=& \hat{S}_{a\,ij} - \frac{\mom{a}{i}(nS_a)_j}{m_a - nP_a} + \frac{\mom{a}{j}(nS_a)_i}{m_a - nP_a}\,.\label{eq:covtoNW}
\end{eqnarray}
Now one has to go from the Newton-Wigner spin tensor in a coordinate basis $S_{a\,ij}$ to the canonical spin tensor
in a triad basis $\spin{a}{i}{j}$ via $\hat{S}_{a\,ij} = \spin{a}{m}{n} e_{i(m)} e_{j(n)}$. 
This canonical spin tensor can be related to the spin vector $\vspin{a}$ via $\spin{a}{i}{j} = 
\epsilon_{ijk} \hat{S}_{a\,(k)}$ or $\hat{S}_{a\,(i)} = \tfrac{1}{2}\epsilon_{ijk} \spin{a}{j}{k}$. 

The transformation going from $\vct{S}_a$ to $\vspin{a}$ using (\ref{eq:31stilde}), (\ref{eq:covtoNW}), the basis transformation, and the relation between spin tensor and spin vector can be compared with $(H^{-1})^{ij}$ calculated perturbatively from $H^{ij}$ mentioned above.
From this calculation one can see that there is a deviation from the canonical spin used in \cite{Damour:Jaranowski:Schafer:2008:1}
and $\vspin{a}$ used here of the form
\begin{eqnarray}
 \hat{S}_{a\,i}^{\text{DJS}} &=& \hat{S}_{a\,(i)} + \frac{1}{c^6} \biggl\{
	\frac{1}{8 m_a^2}\biggl(
		\htt_{(4)ij}\mom{a}{j}(\vmom{a}\vspin{a})\label{eq:sdjsrelateds} \nonumber\\
		&&-\mom{a}{i}\htt_{(4)mn}\mom{a}{m}\hat{S}_{a\,(n)}
		\biggr)
	\biggr\}\,,\\
\nonumber
\end{eqnarray}
where the appearing field variables on the right-hand side have to be evaluated and regularized at 
the position $\vx{a}$ of the $a$th object.
This deviation is three post-Newtonian orders after the leading order.
Note that the lengths of the spins have to be equal (which one can see from (\ref{eq:sdjsrelateds}) is fulfilled 
at the order considered), since $(\hat{\vct{S}}_{a}^{\text{DJS}})^2 = S^2_a = \vspin{a}^2$ with 
$2 S_a^2 = S^{\mu\nu}_a S_{a\mu\nu}$.
The deviation in (\ref{eq:sdjsrelateds}) is beyond the post-Newtonian order considered in the present paper,
thus showing why the Hamiltonians calculated in the ADM formalism are exactly identical to the Hamiltonians 
calculated via spin precession frequency.
}

%% file: refs.bbl
%